\documentclass[twocolumn,amsmath,amssymb,aps,pre]{revtex4}

\usepackage{graphicx}
\usepackage{dcolumn}
\usepackage{bm}
\usepackage{color}

\begin{document}

\title[Electrostatic energy of solid binary ionic mixtures]{Electrostatic energy of solid binary ionic mixtures}

\author{A. A. Kozhberov}
\email{kozhberov@gmail.com}

\affiliation{Ioffe Institute, Politekhnicheskaya 26, Saint Petersburg, 194021, Russia}%

\date{\today}

\begin{abstract}
We study the electrostatic energy of binary ionic mixtures (BIMs) in the form of Coulomb crystals with the main focus on ordered crystals. We consider 15 different binary bcc-like lattices, accurately calculate their electrostatic energies and approximate them by a unified equation. These results extend those available in the literature. A detailed comparison with selected previous results is made, particularly, using previous calculations in the linear mixing rule approximation. The case of disordered BIMs is also outlined. The results are expected to be useful for exploring multi-component Coulomb systems in compact stars, laboratory plasmas, and technological applications.

\begin{description}
\item[PACS numbers]
52.27.Cm, 52.27.Lw, 64.10.+h, 97.60.Jd
\end{description}
\end{abstract}

\maketitle

\section{Introduction}
Coulomb crystals appear in Coulomb plasmas of ions interacting via Coulomb forces and immersed in the neutralizing background of electrons. Such plasmas are fascinating systems due to the long-range nature of Coulomb interaction. The ions can be strongly non-ideal, liquid or solidified. The physics of Coulomb plasma is important for the astrophysics of compact stars, plasma physics, and it has a number of technological applications.

Under the compact stars we mean white dwarfs and neutron stars. Any white dwarf has a bulky and massive core composed of Coulomb plasma of ions (fully ionized atomic nuclei) on the background of degenerate electrons. A neutron star possesses a massive liquid core of nuclear matter surrounded by a light and thin envelope (around $1\%$ by mass) that is usually viewed as a Coulomb plasma of atomic nuclei and degenerate electrons, with an additional background of free degenerate neutrons at densities higher than the neutron drip density, $\sim 4 \times 10^{11}$ g~cm$^{-3}$.  

Here we study Coulomb crystals of point-like ions on uniform background of electrons. The idea of Coulomb crystals was suggested by Wigner \cite{W34} in 1934 who actually invented the crystal of electrons on neutralizing background of ion jelly. Since then Coulomb crystals have been used in various areas of physics (e.g., Ref. \cite{C24}): in solid state physics (e.g., Ref. \cite{F36,BH54,Kit}), non-ideal plasma physics  (e.g., Refs. \cite{F05,B08,F16,EFF17}), and astrophysics of the solar system (e.g., Ref. \cite{FM19}), for interpreting different experiments  (e.g., Ref. \cite{WC74}). A similar two-dimensional Winger crystal has been recently directly observed \cite{L21,T24}.

Needless to say, the Coulomb crystals are important ingredients of compact stars (e.g., Refs. \cite{vH68,HPY07,A10,vH19}). Modern observations and theoretical models strongly suggest the presence of several types of ions simultaneously in the interiors of these stars (e.g., Refs. \cite{MC10,SGC23,B22,JSB21}), but this task is not fully explored and requires new efforts. Molecular dynamics studies of such systems have mainly been performed for some specific mixtures, and they still do not give a general picture (e.g., Refs. \cite{HB09,RR16,CC18,C20}). Even for the simplest case of Coulomb multi-component systems as binary ionic mixtures (BIMs), composed of two types of ions, the theory is far from being complete.

Unlike BIM liquids, the theory of which has been greatly developed in recent years \cite{B19,BY19,BC22}, the state of the theory of BIM solids is much less satisfactory. One of the problems concerns electrostatic energy. The most widely used equation for the electrostatic energy of BIM solids was derived in Ref. \cite{O93}. It was constructed from modest Monte Carlo simulations of only several binary disordered (``random bcc'') Coulomb crystals. Interpretations of observations based on those results (e.g., see \cite{PC13} and references in this paper) are becoming questionable. 

Moreover, it is well known that the electrostatic energy of strongly ordered Coulomb crystals can be smaller than the energy of disordered ones, which attracts new attention to ordered systems. Unfortunately, it is a problem to study ideal crystals with molecular dynamics simulations starting from randomly distributed ions \cite{HB09}. Instead, it is interesting to explore the electrostatic energy of strongly ordered BIM solids using analytic methods.

The electrostatic energy of BIM solids is important for astrophysical applications. It affects nuclear reaction rates and the chemical composition of the neutron star crust and the white dwarf core, directly and through the crystallization process (e.g., Refs. \cite{CF16,HPY07,DG09,FR20,SGC23}). Even for correct interpretations of gravitational wave events from neutron star mergers, it is worth knowing what part of the crust is in the solid state (e.g., Ref. \cite{UCB00}).

In the theory of white dwarfs, the electrostatic energy is crucial for establishing the conditions in structural states of carbon-oxygen mixtures. About 1$\%$ changes in the energy can shift the melting point of the mixture by more than 20$\%$, or even almost completely distort the phase diagram for some compositions (as detailed in Ref.\ \cite{B22}). It seems that the crystallization of mixtures is directly related to the so-called Q branch anomaly (e.g., Refs. \cite{CCH19,TF19,BDS21}) of white dwarfs on the Hertzsprung-Russell diagram, which was firmly established from Gaia observations. It is associated with high-mass white dwarfs but contradicts all generally accepted white dwarf cooling tracks \cite{G18}. This problem is not solved yet. It seems to us that the most reasonable explanation would be to introduce cooling delay of Q-branch white dwarfs during the crystallization and associated separation of ions. However, for simulating Q-branch cooling one needs to elaborate phase diagrams which requires more precise electrostatic energies than those available at the moment. 

It is the aim of this paper to obtain a universal expression for the electrostatic energy of strongly ordered BIM solids and  coping with new challenges in astrophysics \cite{SBT22}.

\section{Ordered BIM solids}
The type of crystal lattice formed in Coulomb plasmas of ions is not certain because the difference between electrostatic energies of crystals of various types is small  (e.g., Ref. \cite{K20}). We adopt the standard assumption that ordered BIM solids possess a body-centered cubic (bcc, CsCl) structure that has been widely discussed in the literature (see, e.g., Refs. \cite{D71,O93,KB12,KB15,CF16,K20}). Note that the bcc lattice has the lowest electrostatic energy among all lattices formed by identical ions in a uniform background of electrons (in one-component plasma of ions, OCP) \cite{KP21}. 

Our ordered binary bcc-like lattice has the same unit cell as the OCP bcc lattice but the ion in the center has a charge number ($Z_2$) different from that of other ions ($Z_1$). Therefore, for the CsCl lattice we have $x\equiv N_2/N=1/2$, where $N_i$ is the total number of ions with the charge number $Z_i$, and $N=N_1+N_2$ is the total number of ions.

The binary bcc-like lattices with $x \neq 1/2$ have been studied rather rarely \cite{O93,INI01,II03}. Our goal is to elaborate these studies. For describing all lattices of our interest, we explore a more complicated ordered five-component bcc lattice. Its unit cell represents eight unit cells of an OCP bcc lattice put into one large cube (see Fig. 2 in Ref.\ \cite{K24}). This lattice can be viewed as a simple cubic lattice with 16 basis vectors: 
$\boldsymbol{\chi}_1=0$, $\boldsymbol{\chi}_2=0.5a_{\rm l}(1,1,1)$, $\boldsymbol{\chi}_3=0.5a_{\rm l}(1,0,0)$, $\boldsymbol{\chi}_4=0.5a_{\rm l}(0,1,0)$, $\boldsymbol{\chi}_5=0.5a_{\rm l}(0,0,1)$,
 $\boldsymbol{\chi}_6=0.5a_{\rm l}(1,1,0)$, $\boldsymbol{\chi}_7=0.5a_{\rm l}(1,0,1)$, $\boldsymbol{\chi}_8=0.5a_{\rm l}(0,1,1)$, $\boldsymbol{\chi}_9=0.25a_{\rm l}(1,1,1)$, $\boldsymbol{\chi}_{10}=0.25a_{\rm l}(3,1,1)$, $\boldsymbol{\chi}_{11}=0.25a_{\rm l}(1,3,1)$, $\boldsymbol{\chi}_{12}=0.25a_{\rm l}(1,1,3)$, $\boldsymbol{\chi}_{13}=0.25a_{\rm l}(3,3,1)$, $\boldsymbol{\chi}_{14}=0.25a_{\rm l}(3,1,3)$, $\boldsymbol{\chi}_{15}=0.25a_{\rm l}(1,3,3)$, and $\boldsymbol{\chi}_{16}=0.25a_{\rm l}(3,3,3)$, where $a_{\rm l}$ is a  lattice constant. These vectors correspond to ions with charge numbers: $Z_{\rm a}$, $Z_{\rm b}$, $Z_{\rm c}$, $Z_{\rm c}$, $Z_{\rm c}$, $Z_{\rm d}$, $Z_{\rm d}$, $Z_{\rm d}$, $Z_{\rm f}$, $Z_{\rm f}$, $Z_{\rm f}$, $Z_{\rm f}$, $Z_{\rm f}$, $Z_{\rm f}$, $Z_{\rm f}$, $Z_{\rm f}$, respectively.

If all ions are fixed at their equilibrium positions, the potential energy of any strongly ordered multi-component Coulomb crystal can be written as
\begin{eqnarray}
U_{\rm M}&=&\frac{1}{2}\sideset{}{'}\sum_{lpl'p'}
\frac{Z_{p}Z_{p'}e^2}{|\textbf{R}_{l}-\textbf{R}_{l'}+\boldsymbol{\chi}_p-
\boldsymbol{\chi}_{p'}|} \\
&-&n_e \sum_{lp} \int \textrm{d} \textbf{r}\,
\frac{Z_{p}e^2}{|\textbf{R}_{l}+\boldsymbol{\chi}_p-\textbf{r}|}
+\frac{n_e^2 e^2}{2}\int\int\,\frac{\textrm{d}\textbf{r}\,
\textrm{d}\textbf{r}'}{|\textbf{r}-\textbf{r}'|}~,  \nonumber 
\label{Upot}
\end{eqnarray}
where $\textbf{R}_l$ is a lattice vector. Sums over $l$ and $l'$ are extended to infinity,
sums over $p$ and $p'$ are over all ions in the elementary cell. In the first sum the terms with $l=l'$ and $p=p'$ should be omitted; $n_e$ is the electron number density. The second and third terms remove divergence in the first sum.

An expression for the Madelung energy of an ordered multi-component Coulomb crystal was derived in our Refs.\ \cite{KB12,KB15}. Using the Ewald transformation \cite{E17,E21} we obtained
\begin{eqnarray}
U_\textrm{M} &=& N\frac{e^{2}}{a}\xi~,
\nonumber \\
\xi&=&\frac{a}{2N_\textrm{cell}}\sum_{lpp'}{Z}_{p}{Z}_{p'}
  \left(1-\delta_{pp'}\delta_{\textbf{R}_l0}\right) \frac{\textrm{erfc}
  \left(A Y_{lpp'}\right)}{Y_{lpp'}}
\nonumber \\
&-&\frac{Aa}{N_\textrm{cell}\sqrt{\pi}}\sum_p Z^2_p
  -\frac{3}{8N_\textrm{cell}^2A^2a^2}\sum_{pp'}Z_{p}Z_{p'}
\nonumber \\
&+&\frac{3}{2N_\textrm{cell}^2 a^2}\sum_{mpp'}{Z}_{p}{Z}_{p'}
    (1-\delta_{\textbf{G}_m0})
\nonumber \\
&\times& \frac{1}{G_m^2}\exp\left[-\frac{G_m^2}{4A^2}+
     i\textbf{G}_m(\boldsymbol{\chi}_p
      -\boldsymbol{\chi}_{p'})\right]~,
\label{Mad_m}
\end{eqnarray}
where ${\textbf{Y}}_{lpp'}={\textbf{R}}_l+\boldsymbol{\chi}_p-\boldsymbol{\chi}_{p'}$, $\textbf{G}_m$ is a reciprocal lattice vector, $N_\textrm{cell}$ is the total number of ions in the elementary cell, $a\equiv\left(4\pi n /3\right)^{-1/3}$ is the mean ion sphere radius,  $n$ is the total number density of ions, and ${\rm erfc}(x)$ is the complementary error function. 
The parameter $A$ is free. It should be chosen in such a way that sums over ${\bf R}_l$ and ${\bf G}_m$ converge equally rapidly. $\xi$ is a dimensionless Madelung constant that depends only on the lattice type and ion charge numbers. For our five-component bcc lattice, it can be written as 
\begin{eqnarray}
-\xi&=&0.0218282388[Z_{\rm a}^2+Z_{\rm b}^2+16 Z_{\rm f}^2] \nonumber \\
&+&0.0123391008\left[Z_{\rm a}Z_{\rm b}+2 Z_{\rm a}Z_{\rm f} \right. \nonumber \\
&+&\left.2Z_{\rm b}Z_{\rm f}+6Z_{\rm c}Z_{\rm f}+6Z_{\rm d}Z_{\rm f}\right] \nonumber \\
&+&0.0268891490[Z_{\rm a}Z_{\rm c}+Z_{\rm b} Z_{\rm d}] \nonumber \\
&+&0.0044282278[Z_{\rm b}Z_{\rm c}+Z_{\rm a}Z_{\rm d}] \nonumber \\
&+&0.0923738654[Z_{\rm c}^2+Z_{\rm d}^2]  \nonumber \\
&+&0.0458737581Z_{\rm c}Z_{\rm d}~.
\label{Mad5}
\end{eqnarray}
This equation allows one to compute electrostatic energies of many cubic lattices, which consist of less than five different types of ions. For instance, it can be used for OCP bcc, face-centered cubic (fcc) and simple cubic (sc) lattices; for multi-component fluorite, perovskite, and Dyson lattices (see Ref.\ \cite{K20} for review), and for many others that have not been considered before. 

\begin{table*}
\caption{\label{tab:1} Values of $\xi_{11} (x)$, $\xi_{12}(x)$ and $\xi_{22}(x)$ in the Madelung energy (\ref{MadU2}) for different binary bcc-like lattices}
\begin{ruledtabular}
\begin{tabular}{cccccc}
$x$ & $Z_1$ & $Z_2$ & $-\xi_{11} (x)$ & $-\xi_{12} (x)$ & $-\xi_{22} (x)$  \\
\hline
1/16 & $Z_{\rm b}$, $Z_{\rm c}$, $Z_{\rm d}$, $Z_{\rm f}$ & $Z_{\rm a}$ &
0.8057663375 & 0.0683346793 & 0.0218282388 \\
1/8 & $Z_{\rm c}$, $Z_{\rm d}$, $Z_{\rm f}$ & $Z_{\rm a}$, $Z_{\rm b}$ &
0.7279425202 & 0.1119911570 & 0.0559955785 \\
3/16 & $Z_{\rm a}$, $Z_{\rm b}$, $Z_{\rm d}$, $Z_{\rm f}$ & $Z_{\rm c}$ &
0.6523296502 & 0.1512257400 & 0.0923738654 \\
1/4 & $Z_{\rm b}$, $Z_{\rm d}$, $Z_{\rm f}$ & $Z_{\rm a}$, $Z_{\rm c}$ & 
0.5890558811 & 0.1657821214 & 0.1410912532 \\
5/16 & $Z_{\rm d}$, $Z_{\rm f}$ & $Z_{\rm a}$, $Z_{\rm b}$, $Z_{\rm c}$ &
0.5156602916 & 0.2005821434 & 0.1796868207 \\
3/8 & $Z_{\rm a}$, $Z_{\rm b}$, $Z_{\rm f}$ & $Z_{\rm c}$, $Z_{\rm d}$ &
0.4546038029 & 0.2107039638 & 0.2306214890 \\
7/16 & $Z_{\rm b}$, $Z_{\rm f}$ & $Z_{\rm a}$, $Z_{\rm c}$, $Z_{\rm d}$ &
0.3957582616 & 0.2164038895 & 0.2837671046 \\
1/2 & $Z_{\rm a}$, $Z_{\rm b}$, $Z_{\rm c}$, $Z_{\rm d}$ & $Z_{\rm f}$ &
0.3492518211 & 0.1974256136 & 0.3492518211 \\
\end{tabular}
\end{ruledtabular}
\end{table*}

Here we use Eq.\ (\ref{Mad5}) for calculating the energies of all allowable BIMs. In particular, we can calculate electrostatic energies of binary bcc-like lattices with eight different $x \leq 1/2$ by setting $Z_{\rm a}$, $Z_{\rm b}$, $Z_{\rm c}$, $Z_{\rm d}$ and $Z_{\rm f}$ equal to $Z_1$ or $Z_2$. These energies can be written in the familiar form 
\begin{eqnarray}
U_{\rm 2M}&=&N\frac{Z_1^2e^{2}}{a}\xi_{\rm bin}(x,\alpha)~, \label{MadU2} \\
\xi_{\rm bin}(x,\alpha)&=&\xi_{11}(x)+\xi_{12}(x)\alpha+\xi_{22}(x)\alpha^2~, \label{Mad2}
\end{eqnarray}
where $\alpha\equiv Z_2/Z_1$.  The ion charges are chosen in such a way for the Madelung energy to be minimal at a given $x$ (see Tab.\ \ref{tab:1}). The parameters $\xi_{11}(x)$, $\xi_{12}(x)$ and $\xi_{22}(x)$ are listed in Tab.\ \ref{tab:1}. The sum $\xi_{\rm bcc}=\xi_{11}(x)+\xi_{12}(x)+\xi_{22}(x)$ is constant, $\xi_{\rm bcc}=-0.8959292557$ at any $x$, which is the Madelung constant of the OCP bcc lattice. As follows from Tab.\ \ref{tab:1}, $\xi_{\rm bin}(1/8,\alpha)=\xi_{\rm bcc} [\alpha^2+2\alpha+13]/16$ and $\xi_{22} (7/16)=13\,\xi_{22} (1/16)$.

It is remarkable that at any $x$ we have
\begin{equation}
\xi_{12}(x)-2 x \xi_{\rm bcc} +2 \xi_{22} (x)=0~.
\label{eq5}
\end{equation}
This expression allows us to rewrite Eq.\ (\ref{MadU2}) as 
\begin{equation}
\frac{U_{\rm 2M}}{N}=\frac{e^{2}}{a}\left[\xi_{\rm bcc} Z^2 +\frac{\xi_{22}(x)- \xi_{\rm bcc} x^2}{(1-x)^2} (Z-Z_2)^2\right]~,
\label{midZ}
\end{equation}
where $Z\equiv(1-x)Z_1+xZ_2$ is the average charge number. 

The BIM energy stays constant under
simultaneous replacements $x \leftrightarrow (1-x)$ and $Z_1 \leftrightarrow Z_2$. This leads to $\xi_{11}(1-x)=\xi_{22}(x)$ and $\xi_{12}(x)=\xi_{12}(1-x)$ owing to which Eq.\ (\ref{Mad2}) can be rewritten as 
\begin{eqnarray}
\xi_{\rm bin}(x,\alpha)&=&\xi_{22}(1-x)+\xi_{22}(x)\alpha^2 \nonumber \\
&+&\left[\xi_{\rm bcc}-\xi_{22}(1-x)-\xi_{22}(x)\right]\alpha~. \label{fit0}
\end{eqnarray}
This equation is valid for any ordered and disordered binary lattices, because only pair interactions are taken into account and the overall form of the $\alpha$ dependence should be the same. 

It is useful to approximate $\xi_{22}(x)$ as a function of $0\leq x \leq1$ disregarding Eq.\ (\ref{eq5}). Unfortunately, this cannot be done with a suitable accuracy, mainly because of the special points $x=3/4$, $x=1/2$ and $x=1/4$ with higher-group symmetry. 

At $x=1/2$ the parameter $\xi_{\rm bin}(1/2,\alpha)$ is the Madelung constant of the CsCl lattice. The details are explained in Refs. \cite{KB15,K20}, where CsCl lattice is called the sc2 lattice.

The point $x=1/4$ is special, and the same is true for $x=3/4$. Among all lattices with $x=1/4$ the binary fcc (fccb) lattice was  studied \cite{KB12,KB15,K20} most attentively. Its Madelung constant is $\xi_{\rm fccb}=-0.5865374846-0.1707354535\alpha-0.138600677\alpha^2$, while $\xi_{\rm fccb}=\xi_{\rm fcc}=-0.8958736151$ at $\alpha=1$. 

If $\xi_{\rm fcc}>\xi_{\rm bcc}$, the bcc lattice has Madelung energy lower than that of the fcc lattice. The situation remains the same at any other $\alpha$ and $\xi_{\rm fccb}>\xi_{\rm bin}(1/4,\alpha)$.  Therefore, the Madelung energy of the fccb lattice is always higher than that of the bcc one. This result can be obtained from Fig.\ 2 in Ref.\ \cite{INI01}. Unfortunately, the exact energy value not given there. 

Consequently, a binary bcc-like lattice has the lowest electrostatic energy not only at $x=1/2$, but also at $x=1/4$. Moreover, since the bcc lattice has the lowest Madelung constant among all lattices at $\alpha=1$, the bcc-like lattices should possess the lowest Madelung energy at least at $|\alpha-1| \ll 1$ for any $x$.

Disregarding the special points $x=1/4$, $x=1/2$ and $x=3/4$, the data for all other $0<x<1$ can be approximated as
\begin{equation}
\xi_{22}(x)=-0.810510\, x^{1.3} \frac{1+0.0525078\, x^{6.63}}{1-0.047840 x}~. \label{fit}
\end{equation}

This approximation looks simple and contains only four fit parameters, because we fix $\xi_{22}(1)=-0.895929 \approx \xi_{\rm bcc}$ and $\xi(0)=0$. Then 12 values from Tab.\ \ref{tab:1} need to be approximated, but they, in turn, are obtained as a linear combination of five independent parameters in the original Eq.\ (\ref{Mad5}) for the electrostatic energy of a five-component lattice. 

The average approximation error is $0.5\%$, and the maximum error is $2.5\%$. It seems problematic to describe electrostatic energies of several discrete ordered structures by one continuous function with  higher accuracy. As shown in the next section, Eq.\ (\ref{fit}) is mostly quite accurate. The accuracy of our approximation can be estimated using Eq.\ (\ref{eq5}). By substituting  Eq.\ (\ref{fit}) into the left-hand side of  Eq.\ (\ref{eq5}) we see that it varies from $-0.00045$ to 0.00045 at $0.1 \leq x \leq 0.9$. This is quite satisfactory, especially if we note that $-0.758 \lesssim \xi_{22}(x) \lesssim -0.0408$ at the same $x$.

The number of ion types in the selected unit shell can be increased to eight by placing new ones into the lattice sites with basis vectors $\boldsymbol{\chi}_{9-16}$. However, such a lattice will not give anything new for our problem. None of the new binary lattices have an energy lower than that of those considered earlier. On the other hand, both, our five-component and eight-component lattices are insufficient for a systematic analysis of Madelung energies of three-component Coulomb crystals. For such an analysis, one should study crystals with a larger unit cell.

Let us add a few words about other assumptions which should be fulfilled. Firstly, we assume a uniform electron background. This is a good approximation for the main part of  a neutron star crust and for a white dwarf interior. 
In Ref.\ \cite{KP21} we studied the effect of polarized background 
of degenerate electrons on electrostatic properties of OCP. It turns  out that if the relativistic parameter 
of degenerate electrons obeys $x_{\rm r} \approx 0.01\,(\rho\, Z/A)^{1/3} <1$, the polarization is important (here $\rho$ is the mass density in units of ${\rm g}/{\rm cm}^3$, and $A$ is the mass number of ions). The structure of degenerat stars at such $x_{\rm r}$ is complicated.

Secondly, in this paper we study the electrostatic energy ignoring zero-point vibrations and any temperature effects. For ordered solid BIMs, these effects have never been studied in detail. In Ref.\ \cite{KB15}  we calculated the zero-point energy and thermal effects in the harmonic-lattice approximation for the bcc-like lattice with $x=1/2$. At some temperatures and ionic compositions, they appeared to be far from the ``standard'' expectations. For other $x$ (see the next section), a similar study is planned to be carried out later.

\section{Comparing with linear mixing rule}
It is advantageous to study the Madelung energy of multi-component crystals by comparing it with
the Madelung energy calculated using the so-called ``linear mixing rule''.This ``rule'' states that any thermodynamic quantity of a multi-component crystal can be represented as a weighted average of thermodynamic quantities of single-component crystals with the same electron number density as in the multi-component crystal \cite{CA90}. For the electrostatic energy of BIM solids, it reads
\begin{equation}
U_{\rm M}^{{\rm lm}}=N_1\frac{Z^{2}_{1}e^{2}}{a_1}\xi_{\rm bcc}+N_2\frac{Z^{2}_{2}e^{2}}{a_2}\xi_{\rm bcc}~,
\label{E_rac1}
\end{equation}
where the effective ion sphere radius $a_j$ for ions of any type $j$ is introduced for a uniform electron number density $n_e$. Then  $a_j=a_eZ_j^{1/3}$, where $a_e\equiv\left(4\pi{n}_e/3\right)^{-1/3}$ is the electron sphere radius. The advantage is that, at certain $x$ and $\alpha$, $U_{\rm M}^{{\rm lm}}$ corresponds to actual ordered lattices considered in Sec.\ II. This allows us to check our results using previous ``mixing-rule'' calculations. 

Equation (\ref{E_rac1}) can be rewritten as
\begin{equation}
U_{\rm M}^{{\rm lm}}=N\frac{Z^{5/3}_{1}e^{2}}{a_e}\xi_{\rm bcc}\left[1-x+x\alpha^{5/3}\right]~.
\label{E_rac2}
\end{equation}
Then the difference of Madelung energies given by Eqs.\ (\ref{MadU2}) and (\ref{E_rac2}) reads
\begin{eqnarray}
\Delta u &\equiv& \frac{U_{\rm 2M}-U_{\rm M}^{{\rm lm}}}{N Z_1^{5/3}e^{2}/a_e} \label{E_rac3} \\
&=&\frac{\xi_{\rm bin}(x,\alpha)}{\left[1-x+x\alpha\right]^{1/3}}-\xi_{\rm bcc}\left[1-x+x\alpha^{5/3}\right]~. \nonumber
\end{eqnarray}
In the case of absolutely accurate calculations, one would have $\Delta u =0$.

For the CsCl lattice, one has $U_{\rm M}^{\rm lm} \leq U_{\rm 2M}$ at $\alpha < 2.4229$; this result was obtained in Ref.~\cite{D71}. For any other lattice one gets $\Delta u > 0$ at any $\alpha$.

\begin{center}
\begin{table}[ht]
\caption{Differences of Madelung energies  $\Delta u$ given by
Eq. (\ref{E_rac3}) and obtained in Ref.~\cite{O93} for various binary bcc-like lattices at $x=1/2$ or $1/4$. \label{tab:2}}
\centering
\begin{ruledtabular}
\begin{tabular}{ccc}
$\alpha$ & $\Delta u$ obtained from Eq. (\ref{E_rac3}) & $\Delta u$ from Ref.~\cite{O93} \\
\hline
 & $x=1/2$ & \\
\hline
$4/3$ & $-0.0000781$ & $-0.00006$ \\
$5/3$ & $-0.000219$ & $-0.00019$ \\
$2$ & $-0.000268$ & $-0.00024$ \\
$3$ & 0.00127 & 0.00134 \\
$4$ & 0.00668 & 0.00678 \\
\hline
 & $x=1/4$ & \\
\hline
$1/4$ & 0.00903262 & 0.009045 \\
$1/3$ & 0.00649246 & 0.006505 \\
$1/2$ & 0.00308744 & 0.0031 \\
$3/5$ & 0.00180437 & 0.00182 \\
$3/4$ & 0.000621427 & 0.00064 \\
$4/3$ & 0.000731892 & 0.00075 \\
$5/3$ & 0.00240028 & 0.00242 \\
2 & 0.00450838 & 0.00453 \\
3 & 0.0113021 & 0.01134 \\
4 & 0.0171511 & 0.01721 \\
\end{tabular}
\end{ruledtabular}
\end{table}
\end{center}

The ``linear mixing rule'' for BIM solids was  previously used in Ref.~\cite{O93}, where $\Delta u$ was extracted from Monte Carlo simulations. The authors discussed the cases with $x=1/2$ and $x=1/4$ for ordered lattices (called there CsCl and 4${\rm \{fcc\}}$, respectively). Their data for 
those lattices are listed in column 3 of Tab. \ref{tab:2}. The second column presents $\Delta u$ which we calculate from Eq.\ (\ref{E_rac3}) for the same $x$ and $\alpha$. 

\begin{figure}
\center{\includegraphics[width=.95\linewidth]{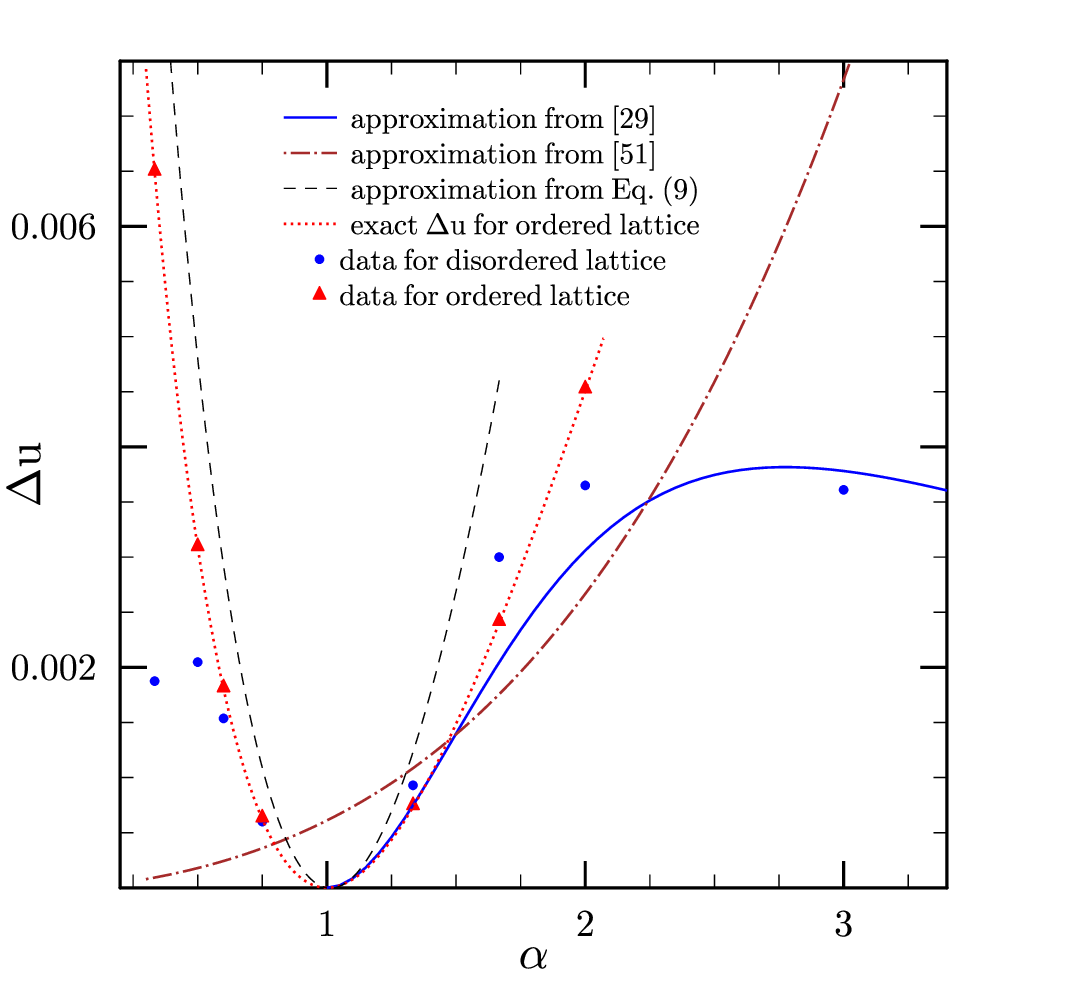}}
\caption{The dependence of $\Delta u$ on $\alpha=Z_2/Z_1$ for $x=1/4$ from Table \ref{tab:2}. The dotted line plots our calculations, while triangles display     the results of Ref.\ \cite{O93}. The dashed line shows our analytic approximation (8). Other lines and symbols refer to disordered crystals discussed in Sec.\ IV.}
\label{fig:4}
\end{figure}

For $x=1/4$, these results are compared in Fig.\ \ref{fig:4}. The dotted line is plotted using Eq.\ (\ref{E_rac3}), while triangles are the points from Ref.\ \cite{O93}. One can see that these results (for ordered lattices) are in a good agreement: relative differences do not exceed a few percent. The maximum difference occurs at $\alpha$ closest to 1, where $\Delta u$ is the smallest and accurate numerical simulations are especially difficult. Note that the authors of Ref.\ \cite{O93} do not present numerical errors for ordered lattices, but present them for disordered lattices. Comparison of columns 2 and 3 in Tab.\ \ref{tab:2} allows us to estimate unaccounted numerical errors in Ref.~\cite{O93}: they are of the same order of magnitude as the accounted one.

\begin{figure}
\center{\includegraphics[width=.95\linewidth]{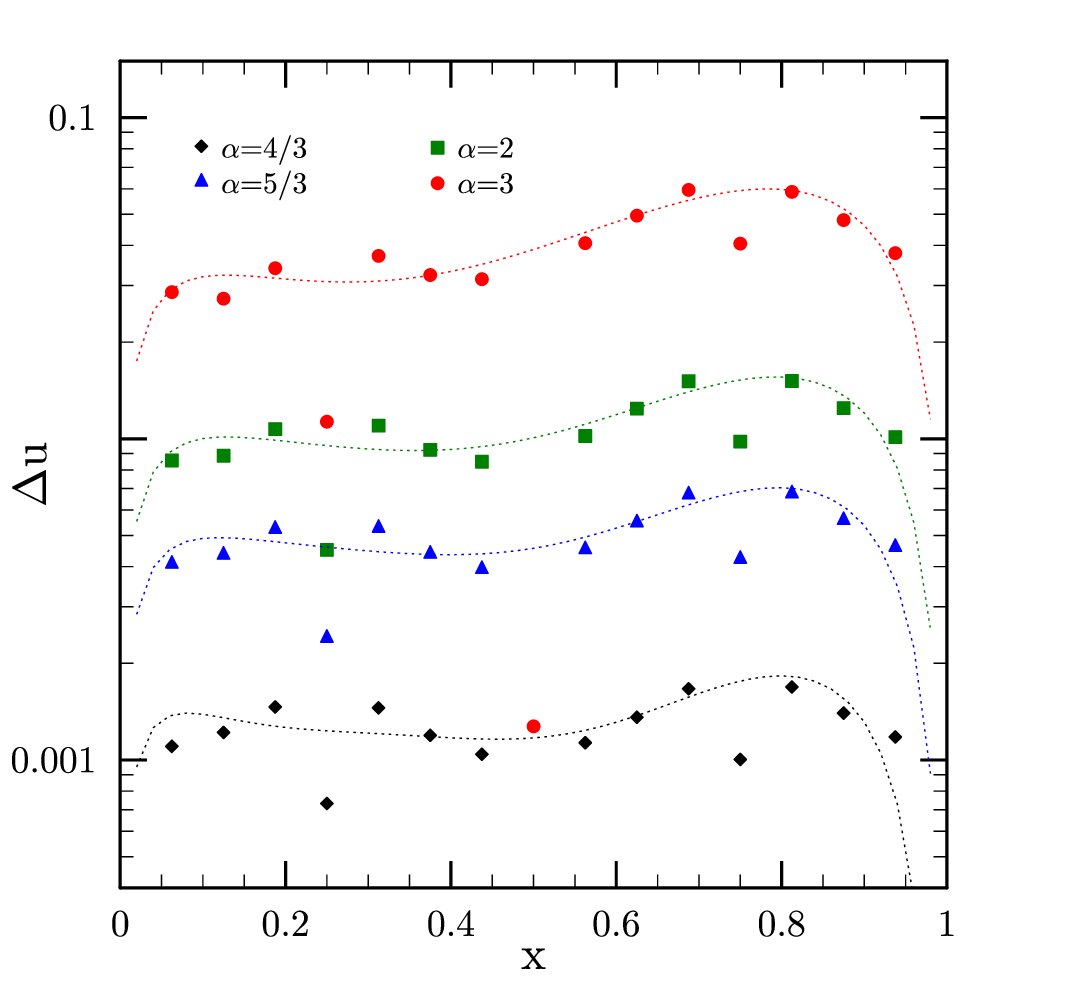}}
\caption{The dependence of $\Delta u$ on $x$ for several values of $\alpha$. Symbols show our results obtained using  Eq.\ (\ref{E_rac3}). They are not distinguishable from those in Ref.\ \cite{O93}. Lines display our analytic approximations, Eqs. (\ref{fit0}) and (\ref{fit}).}
\label{fig:3}
\end{figure}

In Fig. \ref{fig:3} we plot by different symbols our calculations of $\Delta u$ from exact Eq.\ (\ref{E_rac3}) for several $\alpha$ and $x$. The differences between our results and those given in Ref. \cite{O93} in the employed logarithmic scale are not visible. Some points at $x=1/2$ are absent because they are negative (see Tab.~\ref{tab:2}). The dependence of $\Delta u$ on $x$ is non-monotonic. By dashed lines in Fig. \ref{fig:3} we show the dependence of $\Delta u$ on $x$ obtained from analytic approximations (\ref{fit0}) and (\ref{fit}). These lines are plotted for the same $\alpha$ for which exact values are given. One can see that our fit describes $\Delta u$ quite well. The difference between exact and approximate values (for the displayed points) is typically $\sim 10\%$. In the extreme case of $\alpha=4/3$ and $x=1/16$ this difference reaches $34.5\%$, which is maximum for the points displayed in Fig.\ \ref{fig:3}. Equation (\ref{fit}) describes the electrostatic energy of BIMs better than the ``linear mixing rule'' and can be used in applications. A non-monotonic behavior of the points in Fig.\ \ref{fig:3} shows that it is quite problematic to describe $\Delta u$ more accurately.

\section{Disordered BIM solids}
Electrostatic properties of disordered BIM solids were studied in Ref.~\cite{O93} for ``random bcc'' lattices with $x=5/432$, $1/4$, $1/2$ and $3/4$. Based on those data, the authors suggested a fit expression for $\Delta u$ at $1 < \alpha \leq 4$. Simulations for larger $\alpha$ were continued and extended for lattices with $x=1/4$ and $x=1/2$ in Refs.\ \cite{INI01,II03}. In the presence of disorder, $\Delta u$ just measures the difference between the real Madelung energy of 
a given lattice and its Madelung energy calculated in the linear mixing rule approach.  

The results  of Ref.\ \cite{O93} for lattices with $x=5/432$ can be accurately fitted by the
expression $\Delta u/x=-0.2000 + 0.1566 \alpha$. It looks strange because one should have $\Delta u=0$ at $\alpha=1$. We will not discuss this result further. Consequently, we are left with the Monte Carlo results only for those $x$ for which the ordered crystal has structural features. Unfortunately, a direct comparison of the restricted results will not show the overall picture.

As can be concluded from Ref.\ \cite{INI01}, the formation of a disordered lattice at $x=1/2$ can never result in lower Madelung energy. At  $\alpha \lesssim 4.2$ the CsCl lattice possesses the lowest energy, while at $\alpha \gtrsim 4.2$ it is the NaCl (binary simple cubic) lattice which gives the lowest energy. At $\alpha \approx 4.2$ the energies of three lattices become close. On the other hand, the CsCl lattice at $\alpha > 3.596$ is unstable because of instability of some phonon modes, whereas the NaCl lattice is stable at $\alpha > 3.9$ \cite{KB15,K20}. In the small interval of intermediate $\alpha$, none of the known lattices is stable with respect to the growth of phonon modes but a complete analysis of the phonon instability was carried out for a limited number of lattices \cite{K20}.

The situation with deviations from the linear mixing rule at $x=1/4$ is summarized in Fig. \ref{fig:4}, where we plot the dependence of $\Delta u$ on $\alpha$ for different lattices. Solid circles present the data from Ref.\cite{O93} for disordered lattices, and the solid line is their fit. Since the fitting in Ref.\cite{O93} was carried out for lattices with $\alpha >1$, the solid line is broken at $\alpha <1$. The difference between the data and the fit in Ref.\cite{O93} can exceed $20\%$. Our fit is shown by the dashed line. It does not take into account any data at $x=1/4$, but it does not deviate strongly from the whole picture. The larger $\alpha$, the greater the difference from exact results, and  at $\alpha=2$ it reaches $\sim 50\%$.

The fit given in Ref.\ \cite{DS03} is shown by the dot-dashed line in Fig. \ref{fig:4}. At $\alpha=1$ and $x=1/4$ it gives $\xi_{\rm bcc}+\Delta u \approx -0.895319$, which does not correspond to any known one-component lattice. Since no other data (for instance, a lattice type) are given in Ref.\ \cite{DS03}, it is difficult to comment on this result further.

We do not discuss the fit equation (40) presented in Ref.\ \cite{PC13}. New results were not used for its deriving, while this fit does not reproduce old data noticeably better than others.

At small $|\alpha-1|$ the solid circles in Fig.\ \ref{fig:4} lie above triangles (even if we take into account error bars from Ref.\ \cite{O93} and those that were found in the previous section). This means that the formation of an ordered bcc-like lattice is energetically preferable. It has smaller Madelung energy. Note that data for solid circles and solid triangles were obtained in Ref.\ \cite{O93}, and the difference between them was not commented on in that paper. It is obvious that it is very complicated to study a perfect ordered crystal using molecular-dynamic simulations.

At $\alpha \gg 1$ the results of Refs.\ \cite{O93,INI01} are not in line with our theoretical expectations. Equations (\ref{Mad2}) and (\ref{E_rac3}) give $\Delta u \propto \alpha^{5/3}$ for any lattice type, whereas at  $\alpha \gtrsim 2.8$ the solid line in Fig.\ \ref{fig:4} goes down. This may indicate that at such $\alpha$ the lattice is unstable (but inaccuracy of simulations at so high $\alpha$ can also affect the results). The stability of phonon modes for an ordered bcc-like lattice at $x=1/4$ has not been studied in detail.

Generalizing the results for $x=1/2$ and $x=1/4$, we conclude that in a large range of $\alpha$ the Madelung energy of an ordered lattice is lowest.

\section{Conclusions}
We have performed extensive theoretical studies of electrostatic energies of Coulomb crystals composed of point-like ions of different types and a uniform background of electrons. Some crystal types have not been studied before.

In Sec.\ II we calculated the electrostatic energies of 15  ordered BIM crystals and proposed a simple, accurate and unified analytic approximation, Eqs.\ (7) and (\ref{fit}), as a function of two parameters characterizing the ratios of ion numbers and charges, $x=N_2/(N_1+N_2)$ and $\alpha=Z_2/Z_1$. 

In Sec. III we additionally checked the accuracy of our analytic approximation by employing the results of Ref.\ \cite{O93} which were obtained using the so called ``linear mixing rule.'' We re-expressed those results in terms of electrostatic energies of actual BIMs of complicated structure, and found good agreement with our calculations. It is not clear if the proposed analytic fit can be improved further due to the structural features of ordered systems. Their energy does not change monotonically with respect to the parameter $x$ and has peculiarities at $x=1/2$ and $x=1/4$. For these $x$, in Sec.\ IV we compared the energies of ordered and disordered BIM solids and concluded that ordered solids have smaller electrostatic energies.

We expect that our results can be useful for studying phase and structural transitions in complex Coulomb crystals in the envelopes of neutron stars and the cores of white dwarfs, as well
as for interpreting new highest-quality observations of these objects, as outlined in the Introduction.

\begin{acknowledgments}
The author is deeply grateful to D.~G. Yakovlev for help and discussions.
This work was supported by the Russian Science Foundation (grant number 24-12-00320).

\end{acknowledgments}

\end{document}